\documentclass[a4paper,11pt]{article}
\pdfoutput=1 

\usepackage{jcappub} 

\usepackage{verbatim}
\usepackage[utf8]{inputenc}

\usepackage[T1]{fontenc} 

\usepackage{physics}

\makeatletter
\newcommand{\eqnum}{\refstepcounter{equation}\textup{\tagform@{\theequation}}}
\makeatother

\title{Cherenkov Telescope Array sensitivity to branon dark matter models
}

\author[a,b]{A. Aguirre-Santaella,
}
\author[a,b]{V. Gammaldi,}
\author[a,b]{\\ M. A. S\'anchez-Conde,}
\author[c]{D. Nieto}

\affiliation[a]{Instituto de F\'isica Te\'orica UAM-CSIC, Universidad Aut\'onoma de Madrid, \\ C/ Nicol\'as Cabrera, 13-15, 28049 Madrid, Spain}
\affiliation[b]{Departamento de F\'isica Te\'orica, M-15, Universidad Aut\'onoma de Madrid, \\ E-28049 Madrid, Spain}
\affiliation[c]{Grupo de Altas Energ\'ias and IPARCOS, Universidad Complutense de Madrid, \\ Av Complutense s/n, 28040 Madrid, Spain}

\emailAdd{alejandra.aguirre@uam.es}
\emailAdd{viviana.gammaldi@uam.es}
\emailAdd{miguel.sanchezconde@uam.es}
\emailAdd{d.nieto@ucm.es}

\abstract{In the absence of a clear hint of dark matter (DM) signals in the GeV regime so far, heavy, $\mathcal{O}$(TeV) DM candidates are gradually earning more and more attention within the community. Among others, extra-dimensional \textit{brane-world} models may produce thermal DM candidates with masses up to 100 TeV. 
These heavy DM candidates could be detected with the next generation of very-high-energy gamma-ray observatories such as the Cherenkov Telescope Array (CTA). 
In this work, we study the sensitivity of CTA to branon DM via the observation of representative astrophysical DM targets, namely dwarf spheroidal galaxies. In particular, we focus on Draco and Sculptor, two well-known dwarfs visible from the Northern and Southern Hemisphere, respectively. For each of these targets, we simulated 300\,h of CTA observations and studied the sensitivity of both CTA-North and CTA-South to branon annihilations using the latest publicly available instrument response functions and most recent analysis tools. 
We computed annihilation cross section values needed to reach a $5\sigma$ detection as a function of the branon mass. Additionally, in the absence of a predicted DM signal, we obtained $2\sigma$ upper limits on the annihilation cross section. 
These limits lie $1.5-2$ orders of magnitude above the thermal relic cross section value, depending on the considered branon mass. 
Yet, CTA will allow to exclude a significant portion of the brane tension-mass parameter space in the $0.1-60$ TeV branon mass range, and up to tensions of $\sim 10$ TeV. More importantly, CTA will significantly enlarge the region already excluded by AMS and CMS, and will provide valuable complementary information to future SKA radio observations. We conclude that CTA will possess potential to constrain brane-world models and, more in general, TeV DM candidates.}

\begin{document}
\maketitle
\flushbottom

\section{Introduction}

The standard cosmological model states that ordinary matter represents only $\sim$4\% of the total energy content of the Universe~\cite{planck15}. 
Everything else is still unknown: 27\% is best described to be some kind of non-luminous matter, known as dark matter (DM), and the remaining 69\% an enigmatic dark energy. Despite all the evidence accumulated at very diverse spatial scales supporting the existence of a dark component of matter, 
we do not know yet what the DM is made of \cite{2018RvMP...90d5002B}. One of the preferred DM particle candidates are the so-called weakly interacting massive particles (WIMPs),
that arise naturally from well motivated extensions of the Standard Model (SM) of particle physics; see, e.g., \cite{Bertone10,Roszkowski+17} for a review. Three methods have been proposed to find WIMPs: direct production at particle accelerators, direct detection through scattering off a baryonic target, and indirect detection. The latter consists on searching for secondary products of annihilation or decay of DM particles into SM particles. These products, such as photons, carry critical information that might help to elucidate the DM nature~\cite{Porter:2011nv}. 

The gamma-ray astronomy is of particular interest for the indirect detection of WIMPs with masses ranging from MeV to TeV. Indeed, highly energetic photons resulting from WIMP annihilation might be visible to current gamma-ray experiments such as \textit{Fermi}-LAT~\cite{fermiglast}, HAWC~\cite{Smith:2007ni} and imaging atmospheric Cherenkov telescopes (IACTs),
such as H.E.S.S.~\cite{hinton_2004}, MAGIC~\cite{Lorenz:2004ah} or VERITAS~\cite{Weekes:2001pd}. 
IACTs are instruments sensitive to very-high-energy (above 10 GeV) gamma rays, by taking images of the extended air showers that these gamma rays, as well as cosmic rays, produce when they are absorbed in the atmosphere. The information included in those images is then used to derive the properties of the shower progenitor: particle type, energy and arrival direction. Yet, DM has not been univocally detected through this technique as of today (neither by any other), despite all the efforts made in this direction, e.g.~\cite{2016PhR...636....1C, 2018JCAP...03..009A, 2018PDU....22...38A, 2017PhRvD..95h2001A, 2018JCAP...11..037A, 2018PhRvL.120t1101A} and references therein. 

In the near future, new gamma-ray observatories will surely continue the search. Among others, the Cherenkov Telescope Array (CTA)\footnote{\url{https://www.cta-observatory.org/}}~\cite{2011ExA....32..193A, 2013APh....43....3A, 2019scta.book.....C}, an international project that will build the next generation ground-based gamma-ray observatory, will be well positioned to shed further light on the nature of DM. CTA will be around ten times more sensitive than current IACTs in the energy range between few tens of GeV up to 300 TeV. The observatory will consist of two arrays: one located in the Northern Hemisphere (La Palma, Canary Islands, Spain); the other situated in the Southern Hemisphere (Atacama Desert, Chile). The former will be mainly devoted to study extragalactic objects while the latter will possess a better access to the Galactic center.

In this work, we will study CTA sensitivity to a particular type of WIMPs, i.e., \textit{branons}~\cite{Cembranos:2003mr}. Branons generate from the effective theory of extra-dimensional \textit{brane-world} and may represent thermal DM candidates with masses up to 100 TeV; thus they represent good DM candidates for CTA which, as said, is well suited for TeV DM searches. 
In the past, a thermal branon with mass of $\sim 50$ TeV was proposed to explain the cut-off at TeV scale observed by the H.E.S.S. telescope at the Galactic center~\cite{Gammaldi:2019mel, PhysRevD.86.103506, Cembranos:2013fya, Gammaldi:2016uhg}. 

To evaluate the CTA sensitivity to the branon DM model we will use the latest version of the CTA science analysis tools and publicly available instrument response functions. 
We will adopt two astrophysical targets as benchmarks, one for each site, namely the Draco and Sculptor dwarf spheroidal (dSph) galaxies. This kind of objects has been shown to be particularly relevant for DM searches given their high mass-to-light ratios, relative proximity and low level of expected gamma-ray background induced by ordinary astrophysical phenomena \cite{2016PhR...636....1C, 2016ApJ...832L...6W, 2020arXiv200313482R}. In particular, Draco and Sculptor are both very well studied objects and are among the brightest ones in terms of expected DM flux \cite{2011JCAP...12..011S,Bonnivard:2015xpq,2015ApJ...809L...4D}.

The paper is organized as follows. In section~\ref{sec:anibra} we introduce the generalities of indirect DM searches and describe the properties of the gamma-ray spectra generated by branons of different masses. Section~\ref{sec:met} is devoted to CTA's main technical aspects and comprises the details of both the simulated observations and the DM analysis strategy as well (if the reader is not interested in such specialized details, they can skip subsection~\ref{sec:submet}). We show our results in section~\ref{sec:resul}, also placing them into a more general context. Finally, we summarize our main findings in section~\ref{sec:conclu}.

\section{Gamma-ray fluxes from branon annihilation} 
\label{sec:anibra}

\begin{figure}[h]
\begin{center}
\includegraphics[width=\textwidth]{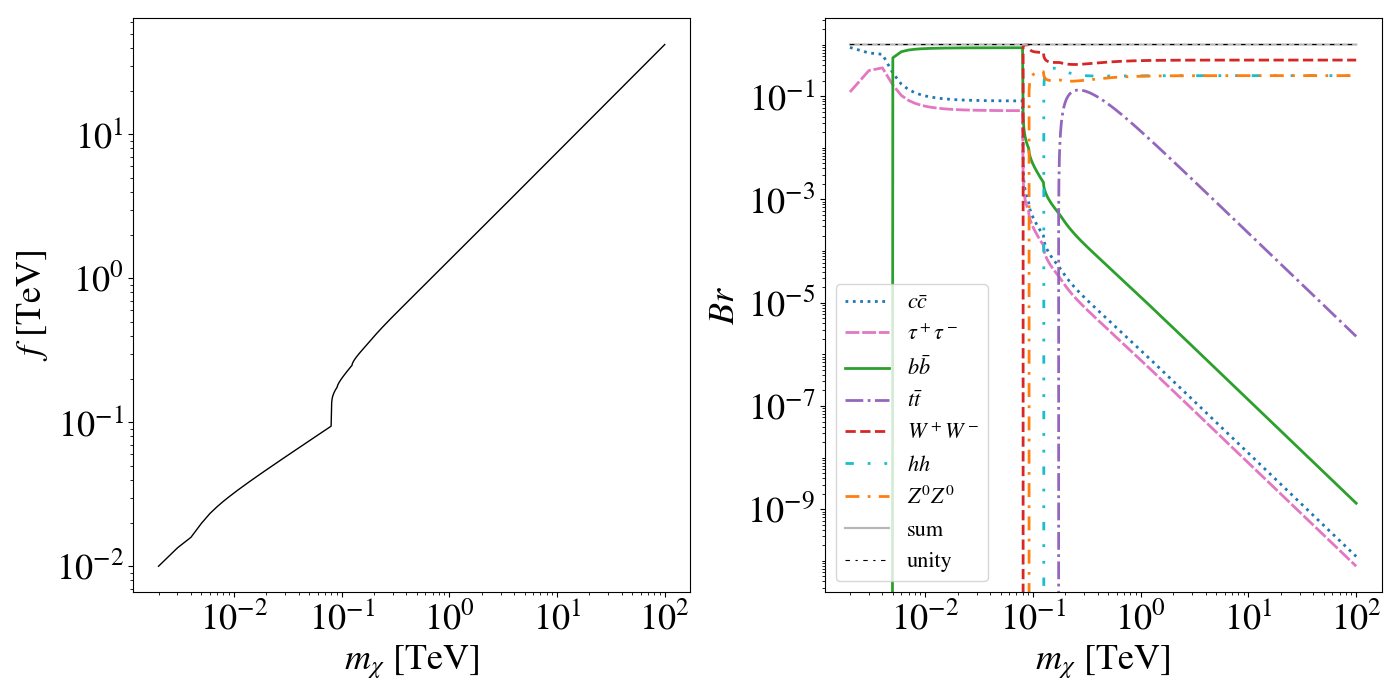}
\caption{Left: tension of the brane (f) as a function of the DM (branon) mass ($m_\chi$) when the branon is considered to be a thermal relic and can account for the total DM content of the Universe. Right: Branching ratios of the most relevant annihilation channels, defined according to eq.~\ref{eq:myeqn2}, for branons with a mass ranging from a few GeV to 100 TeV. The sum of all branching ratios at a given mass is the unity. 
Expressions for $f$ and branching ratios taken from~\cite{Cembranos:2003fu}. 
} 
\label{fig:bryf}
\end{center}
\end{figure}

In the framework of extra-dimensional models, one popular extension of the SM is the so-called brane-world scenario, where SM particles are considered to be bound to a spatial three-dimensional
brane embedded into a higher dimensional $D = 4 + N$ space-time (with $N$ extra dimensions), called the bulk, where gravity is able to propagate~\cite{Cembranos:2003mr, Cembranos:2003fu}. In this scenario, the fundamental scale of gravitation is not the Planck scale, yet it could reach the electroweak scale. On the one hand, this fact allows to solve the hierarchy problem. On the other hand, it allows to model extra-dimensional theories that include the gravitational interaction. 
In this model, branons represent the vibrations of the brane in the direction of the extra-dimensions, i.e. branons acquire mass by a symmetry breaking in those directions. Thus, branes are flexible and this property can be characterized by a tension $f$. In the weakly coupled theory, i.e. $f \gtrsim m_\chi/(4\sqrt{\pi})$ \cite{PhysRevD.73.035008}, where $m_\chi$ is the particle mass, branons couple with the SM particles as the inverse of the fourth-power of the tension of the brane, i.e. they are weakly interacting as $ f^{-4}$ \cite{Cembranos:2003mr}.
 WIMPs as they are, branons represent a viable candidate for DM. In the simplest case of this effective field theory, we only have one extra-dimension, that is, $N=1$ and one branon particle. As WIMPs, branons may then annihilate into e.g. a pair of quarks, a pair of weak bosons, or even a pair of photons, yet the probability for the latter to occur is extremely low \cite{Cembranos:2003mr}. The branching ratio of annihilation into each SM channel depends on the mass of the branons ($m_\chi$) and $f$. In case branons are considered thermal relics and their annihilation cross-section value is the one needed to account for 100\% of the total DM content of the Universe, the tension is a function of the branon mass, and we are left with only one free parameter. This is shown in the left panel of figure~\ref{fig:bryf}. The right panel of the same figure contains the branching ratios for different branon masses. It can be seen that, in our relevant energy band, if $m_\chi$ is below the kinematic threshold for $W^+W^-$ production, the $b\bar{b}$ channel dominates; otherwise, the $W^+W^-$ channel is the most significant one~\cite{PhysRevD.85.043505}. 
 
 Since we need two DM particles to have an annihilation, and each annihilation can produce several photons, the number of potential encounters within DM particles will be proportional to the DM number density, $n_{\chi}$, squared, and the number of generated photons will be proportional to $n_{\chi}^2$ times the velocity-averaged annihilation cross section, $\expval{\sigma v}$, which represents the interaction likelihood~\cite{MC}. With this in mind, the differential annihilation flux will thus be

\begin{equation}
\dfrac{d\phi}{dE} = \displaystyle \int_{l.o.s.} \rho^2 (r)\, dr \, \dfrac{\expval{\sigma v}}{2 \pi m^2_\chi} \sum_i Br_i \dfrac{dN_i}{dE} \label{eq:myeqn2} \end{equation}

Indeed, the astrophysical factor, $J = \displaystyle \int_{l.o.s.} \rho^2 (r)\, dr$, corresponds to the DM density squared, taking into account the integration along the line of sight (l.o.s.) of the DM density distribution $\rho(r)$ in the source. In our work, we will not include J-factor uncertainties and we will simply consider benchmark values in section~\ref{sec:resul}. 
Instead, we will pay particular attention to the factor referring to the photon yield, i.e. the particle physics factor. Notice that $\expval{\sigma v}_\text{th}= 3 \times 10^{-26} \,\text{cm}^{3}\,\text{s}^{-1}$ is the thermally-averaged cross section of the DM particle times the velocity~\cite{2012PhRvD..86b3506S}, and $Br_i=\expval{\sigma v}_\text{i}/\expval{\sigma v}$ is the branching ratio of annihilation channel $i$.

\begin{figure}[h]
\begin{center}
\includegraphics[width=\textwidth]{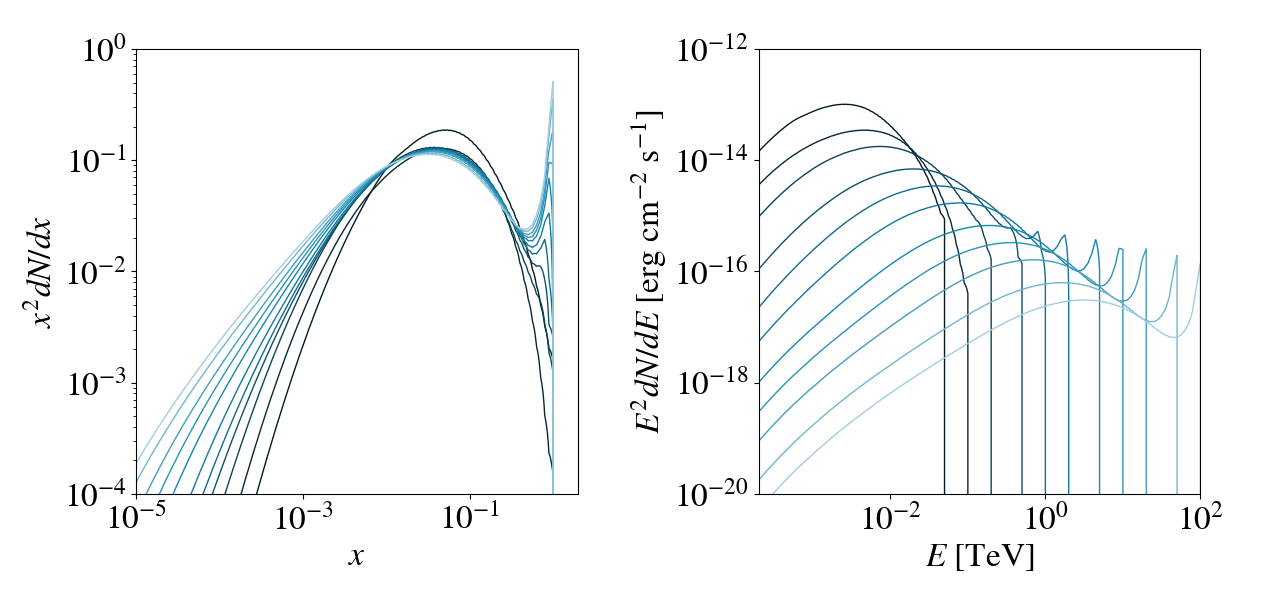}
\caption{Annihilation spectra for different branon masses, from dark to light blue: 0.05, 0.1, 0.2, 0.5, 1, 2, 5, 10, 20, 50 and 100 TeV.
Left: photon yield in terms of the energy normalized to the branon mass, $x =E_\gamma/m_\chi$, times this normalized energy squared. Right: differential flux times the energy squared in terms of the energy for the example case of adopting $J=1.42 \cdot 10^{19}$ GeV$^2$ cm$^{-5}$ (similar to Draco's; see section~\ref{sec:resul}). Photon fluxes obtained with the tools in \cite{Cirelli:2010xx, Ciafaloni:2010ti}.
}
\label{fig:spec}
\end{center}
\end{figure}

Assuming the branon branching ratios of figure~\ref{fig:bryf}, the expected gamma-ray flux from the annihilation of branon DM has a characteristic spectrum, depending on the mass of the particle itself, as we show in figure~\ref{fig:spec}. The energy-normalized spectra 
have roughly the same shape, specially at intermediate values of the dimensionless parameter $x = 2E_\gamma/E_\text{CM}$, where $E_\gamma$ is the energy of the emitted gamma-ray photons and $E_\text{CM}$ is the energy of the center of mass of the simulated event. This variable is simply reduced to $x =E_\gamma/m_\chi$ in the case of annihilating DM~\cite{MC}. We note that the electroweak corrections used in these calculations are model-independent and they have been computed at leading order; ideally, further corrections should be taken into account above several TeV~\cite{Cirelli:2010xx, Ciafaloni:2010ti}. For this reason, although we have also computed the spectrum for 100 TeV branons for illustrative purposes in figure~\ref{fig:spec}, we will stop our calculations at 60 TeV and will show results only up to this energy in the next sections. This particular choice is also motivated by the observations of very high energy gamma-rays, suggesting a $\sim$50 TeV branon, at the Galactic center with HESS~\cite{2013ehep.confE.103C}.

\section{CTA sensitivity}
\label{sec:met}

\subsection{The Cherenkov Telescope Array}

CTA is an international project to build the next generation ground-based gamma-ray observatory, that will be sensitive to gamma rays in the 20 GeV to 300 TeV energy range~\cite{2019scta.book.....C}. CTA is proposed as an open observatory and will consist of two arrays of IACTs, more than 100 telescopes in total, one at the Northern Hemisphere (La Palma, Canary Islands, Spain), with emphasis on the study of extragalactic objects down to the lowest CTA achievable energies, and another array at the Southern Hemisphere (Atacama Desert, Chile), that will cover the full energy range and will mostly concentrate on Galactic sources. CTA will feature three types of IACTs~\cite{2017APh....93...76H}: Large-Sized Telescopes (LSTs)~\cite{2019ICRC...36..653C}, designed to capture the faintest air showers, Medium-Sized Telescopes (MSTs)~\cite{2019ICRC...36..269G}, enlarging the instrument's effective area and contributing to the core energy range of CTA, and Small-Sized Telescopes (SSTs)~\cite{2017ICRC...35..758S}, further expanding the array's footprint and increasing its capability to detect the highest energy events.

The CTA Consortium science program, covering almost half of the available observing time over the first ten years of operation, is made up of individual key science projects that go beyond high-energy astrophysics into cosmology and fundamental physics~\cite{2019scta.book.....C}. These include performing both a galactic and an extra-galactic survey, an in-depth observation of the Galactic center region, studies of star formation, active galactic nuclei, galaxy clusters and transient phenomena. It also includes an ambitious DM program, given CTA's great capabilities for DM searches, i.e. improved sensitivity in the whole energy range, increased field of view and better angular and energy resolutions with respect to the current generation of IACTs. The Galactic center, the Large Magellanic Cloud, dSphs and galaxy clusters are all considered as valuable DM targets for CTA. The total observing time that has been proposed for the DM program in the first three years of operation amounts nearly 1,200$~h$, plus an additional
300$~h$ for the Perseus galaxy cluster.\footnote{Actually, a large fraction of this observation time is not motivated by DM alone and will be used for other science cases as well; this is the case of, e.g., the observations of the Galactic center or the Large Magellanic Cloud.}

\subsection{Analysis pipeline 
}
\label{sec:submet}

In this section, we will describe the simulations of the detection and data analysis procedure of the observations of Draco and Sculptor by the CTA Observatory.

The simulation of the performance of CTA is a multi-stage process (see~\cite{2013APh....43..171B,2017APh....93...76H} for further details). First, the simulation of extended showers in the atmosphere is performed, including the propagation of the Cherenkov photons through the optical system of the telescopes and how these are detected by the fast cameras of the telescopes and converted into digitized pulses. Those pulses are later calibrated to form images of the extended showers. Second, by analyzing images of the same shower as seen by different telescopes, the properties of the original shower can be inferred, which allows for an estimation of the energy and arrival direction of the shower, and for an efficient background suppression. The main products out of this analysis, carried out within the CTA Consortium by the Analysis and Simulation working group, are the instrument response functions (IRFs). 
More specifically, the IRFs provide a relation between the properties of an event as they are measured in the detector and the actual physical properties of the incident particle, i.e., they define the response of the instrument to gamma rays of a given energy and arrival direction --which are inferred assuming that the event is actually a photon-- and they also encapsulate the properties of the distribution of observable events. They thus depend on the energy, exposure time, and particle arrival direction.

Specifically, we have used the \texttt{ctools} software package 
\cite{2016A&A...593A...1K}
to both generate the simulations and run the subsequent data analyses. \texttt{ctools} was conceived to offer easy software tools that could be used to analyze high-level data 
collected by IACTs. The software ope\-rates with reconstructed event lists.
The user feeds \texttt{ctools} with the IRFs, that describe the transformation between physical photon features and measured event features. IRFs also add background, i.e., events that look like photons to the analysis but are actually cosmic rays (electrons and hadrons). 
In our case, we have made use of the publicly available latest CTA IRFs\footnote{\url{https://www.cta-observatory.org/science/cta-performance/}}
(\texttt{prod3b-v2}). The main IRFs are the effective area, the energy resolution, the angular resolution and the background rate.

In the following, we explain the main steps of our analysis pipeline: 

\begin{itemize}
\item Starting from the annihilation spectra, $d\phi/dE$, for a given branon mass, the coordinates of the source and its J-factor, the gamma-rays event list (containing the energy of each simulated photon, arrival time and position) is simulated.

First, we need an input spectrum to generate simulations. There is not yet an available function in \texttt{ctools} that could be used to natively build DM spectra, although it is possible to input generic spectra as tables. Therefore, starting from the so-called PPPC4DMID tables in ref.~\cite{Cirelli:2010xx}, we create our own tables for each branon mass, containing differential flux values as a function of energy, as given by the considered branon model, $d\phi/dE$. To describe the latter, for convenience we introduce a normalization $N_0\, (>0)$ defined such as $N_0 = 1$ for branon models that provide us with the thermal relic cross section value, see eq.~\ref{eq:myeqn2}, and it will be different (larger) for branon models that adopt cross sections other than the thermal; i.e., $N_0$ indicates how far the cross section for the considered branon model, $M$, is from the thermal value:

\begin{equation}
\label{eq:model_equation}
M = N_0 \dfrac{d\phi}{dE} 
\end{equation}

We use the \texttt{ctobssim} tool to simulate event lists for CTA given an input spectrum model. To do so, this tool already accounts for the instrument performance via the inclusion of the IRFs. In addition to both the spectral model and the IRFs, \texttt{ctobssim} requires the sky coordinates of the source, the radius of the simulated region, observation time and desired energy range. The stochastic nature in both the emission of astrophysical gamma-ray photons and the detection of these by CTA are accounted for in the simulations, and independent observations can be produced by setting a seed value for the utilized random number generator. The simulation includes astrophysical photon counts from the source and background event counts from an instrumental background model.

The observation mode is determined by means of the script \texttt{csphagen}, which generates the support files that allow for a subsequent region-based spectral analysis. We chose the wobble observation strategy~\cite{1994APh.....2..137F}, frequently adopted by current IACTs, and the basis for pointing optimization strategies for indirect DM searches~\cite{2019APh...104...84P}. This strategy consists on observing the source with an offset with respect to the center of the field of view, such as not only the source region but also one background region (or several) can be observed at the same time. This allows for a better control of the involved systematics, as both the source and background regions are observed at the same time and under the same observational conditions. Indeed, since the background rates are expected to be approximately radially symmetric in camera coordinates, by wisely choosing the background region(s) this method minimizes the impact of the background rate modeling. Also, the wobble mode reduces the observation time needed to achieve a given sensitivity with respect to other possible observational methods, since no dedicated background observations are necessary.\footnote{We note however that, as it will be explained in this same section, in our specific analysis we will adopt a likelihood method, for which an estimate of background region counts is not necessary and, thus, a wobble strategy would not be strictly necessary either.}

\item We make use of \texttt{ctlike} in order to apply a full likelihood analysis to the event lists for our simulated observations. Note that, in our case, this analysis maximizes the sensitivity to DM-induced $\gamma$-ray signals by implicitly including the spectrum of each considered DM model as well.\footnote{The improvement in sensitivity has been estimated to be about $\sim 1.5 - 2.5$ with respect to the conventional analysis in which no specific DM spectral information is considered, but only a generic power-law index typical of astrophysical sources~\cite{2012JCAP...10..032A}. Ultimately, though, the gain will depend on the precise spectral shape of the DM signal as well as on the potential correction that any trial factors, as derived from the look-elsewhere effect, may introduce in the pre-trials result.}  More precisely, the likelihood analysis will quantify the compatibility of our data with the so-called null hypothesis, i.e., no DM signal in the data.
With \texttt{ctlike}, we compute a test statistic (TS) of the form: 

\begin{equation}
\label{eq:TS_equation}
TS=-2\cdot \textrm{ln}\left[\frac{\mathcal{L}(H_0)}{\mathcal{L}(H_1)}\right]
\end{equation}

\noindent where $\mathcal{L}(H_0)$ and $\mathcal{L}(H_1)$ are the likelihoods under the null (no DM) and alternative (exis\-ting DM source) hypotheses, respectively. By default, \texttt{ctlike} adopts Poisson statistics for the likelihoods' computation. Ultimately, \texttt{ctlike} generates an output file with the values of the parameters maximizing the likelihoods as well as their corresponding errors.

\item As a last step, we obtain values of the annihilation cross section for two cases: i) detection with CTA ($TS=25$); and ii) 95\% c.l. upper limits ($TS=2.71$)~\cite{2011EPJC...71.1554C, 2015APh....62..165C, 2020arXiv200313482R}.

\begin{itemize}
\item[i)] In order to study branon models detection with CTA, we develop the following algorithm. We first run the analysis pipeline described in this section and evaluate the TS value obtained from \texttt{ctlike}. Should this TS differ from 25, on a next iteration we will either increase or decrease the value of $N_0$ in eq.~\ref{eq:model_equation} by means of a multiplicative factor, depending on whether we have got a smaller or larger $TS$, respectively. Effectively, this is analogous to increasing or decreasing $\expval{\sigma v}$ (once the J-factor of the source has been fixed, as it is indeed the case). 
We then repeat the previous steps and the procedure here described until we reach convergence. We allow for a tolerance of $\pm 0.25$ in the obtained TS value. Thus, as soon as a value of $TS = 25 \pm 0.25$ is reached after one of the iterations, we stop the loop and save the obtained value of $\expval{\sigma v}$.

\item[ii)] On the other hand, we obtain 95\% confidence level upper limits to the annihilation cross section. In this case, we use \texttt{ctulimit}, that allows to obtain upper limits to the flux at the desired confidence level for a specific source model~\cite{2016A&A...593A...1K}. Starting from the maximum likelihood model parameters, \texttt{ctulimit} finds the model flux that leads to a decrease of the likelihood corresponding to a given confidence level, using $N_0=1$. Then, in order to get the upper limits to the annihilation cross section, $\expval{\sigma v}^{UL}$, we just need to use the obtained flux upper limit and the following expression, which can be easily derived from eq.~\ref{eq:myeqn2}: 

\begin{equation} \label{eq:ulims}
   \expval{\sigma v}^{UL} = \frac{8\, \pi\, m_\chi^2\, \Phi^{UL} (>E_{min}) }{J\, \int_{E_{min}}^{m_\chi} \frac{d\phi_\gamma}{dE} dE}
\end{equation}

where $\Phi^{UL} (>E_{min})$ is the integrated energy flux obtained using \texttt{ctulimit}, starting at energy $E_{min}$ (that will be shown later in table~\ref{tab:simu}), and $\frac{d\phi_\gamma}{dE}$ is the DM-induced photon spectrum. 

\end{itemize}
\end{itemize}

\begin{figure}[ht]
\begin{center}
\includegraphics[width=.9\textwidth]{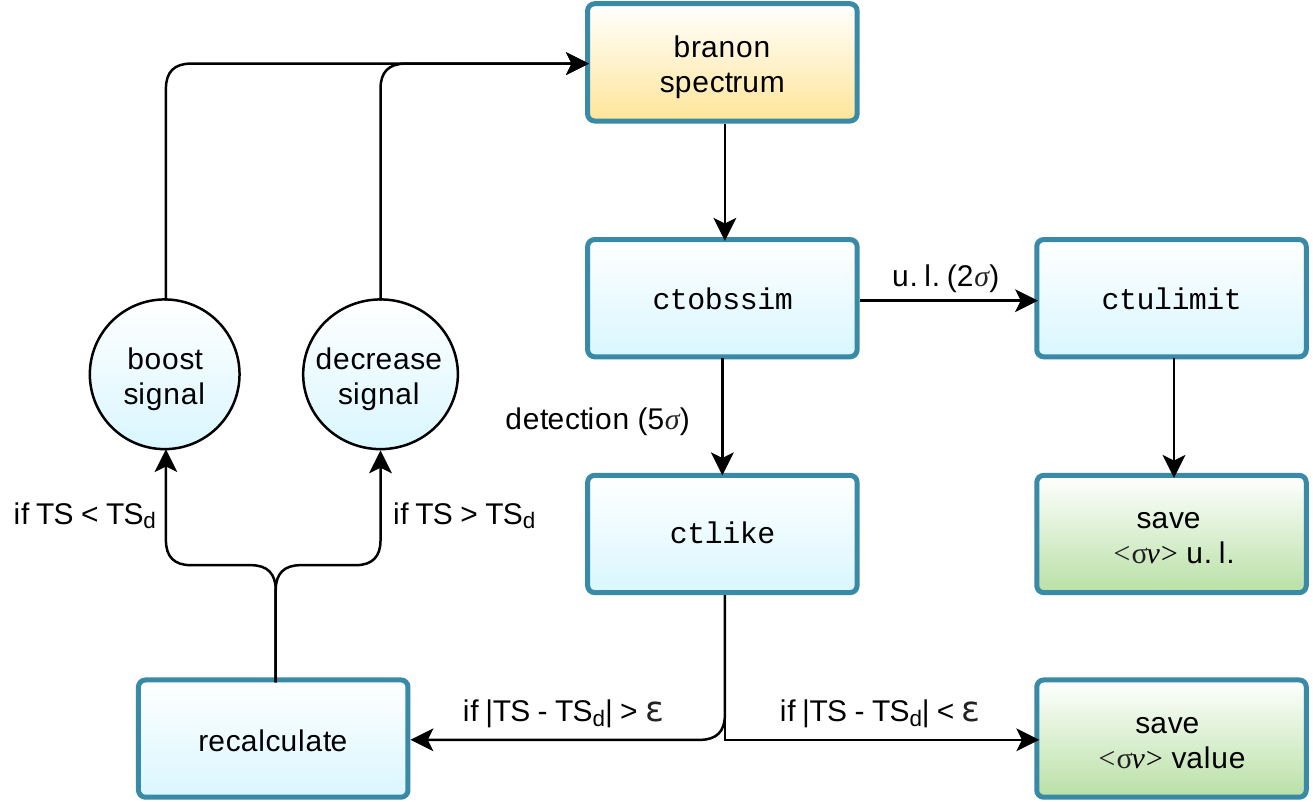}
\caption{Flowchart of the simulation chain, both for the case of studying branon detection and for setting upper limits on the annihilation cross section; see section~\ref{sec:submet} for details. In this chart, $TS_d$ is the TS of detection ($TS_d=25$ for a $5\sigma$ detection case); $\varepsilon$ represents the tolerance used for TS convergence, set to 0.25 in our algorithm.}
\label{fig:flowc}
\end{center}
\end{figure}

Figure~\ref{fig:flowc} shows a flowchart that summarizes the full simulation chain. 

\section{Detection prospects and constraints on branon DM models}\label{sec:resul}

\subsection{Targets}

As mentioned before, we selected two dSphs for our analysis. DSphs have been identified as one of the best targets for gamma-ray DM searches, given their high mass-to-light ratios, their proximity and abundance, and lack of gamma-ray astrophysical backgrounds, e.g.~\cite{2011JCAP...12..011S, dwarf1, 2016PhR...636....1C, 2020arXiv200313482R}. More specifically, we decided to focus on Draco and Sculptor as representative cases of the so-called ``classical'' dwarfs (i.e., those discovered before SDSS data became available in 2006). Classical dSphs have the advantage of having more robust J-factor determinations compared to more recently discovered ones \cite{Bonnivard:2015xpq, 2015ApJ...809L...4D, 2017ApJ...834..110A}, thanks to a better determination of their underlying DM distributions. This is so because they typically possess larger masses and, as such, they have a larger number of star members for which kinematics data is available. The chosen dSph parameters are shown in table~\ref{tab:simu}. Note that we chose one dSph in each hemisphere, this way allowing us to perform a study of the sensitivities of both CTA-South and CTA-North to branons. Both dSphs are assumed to be point-like, which is a fair approximation given the angular size subtended by the bulk of the DM-induced emission in these objects and the CTA PSF.\footnote{Actually, the authors in ref.~\cite{Bonnivard:2015xpq} quote 0.28 and 0.38 degrees as optimal integration angles for annihilation in the case of Draco and Sculptor, respectively, while the CTA PSF, though highly energy-dependent, will be below 0.15 degrees at $\sim 100$ GeV and significantly better than 0.1 degrees at TeV energies, see, e.g.,~\cite{2017APh....93...76H}. Thus, we note that our assumption of these objects being point-like for CTA --though extremely convenient for the analysis, reasonable, and in line with previous works that used IACTs to search for DM in dwarfs-- should be considered as a first approximation only. Indeed, a proper CTA analysis of dwarfs as extended sources, for which new analysis tools may also be necessary, is beyond the scope of this work and will be done elsewhere.}

For each selected target we simulate CTA observations with total exposures of 300 h per dSph. This exposure is the expected amount of time dedicated to observations of dSphs in the first 3 years of scientific operation of CTA~\cite{2019scta.book.....C}. Note that the coordinates in table~\ref{tab:simu}
refer to the target position, which slight differs from the center of the CTA field of view when adopting the wobble strategy for the observation. In our case, the pointing has an offset of 0.5 degrees with respect to the source position~\cite{2019APh...104...84P}. The radius chosen for both the On and Off regions of the wobble observation is 0.2 degrees. 
We have adopted \texttt{North\_average\_z20} and \texttt{South\_average\_z20} as IRF files for Draco and Sculptor, respectively, which implicitly refer to the Northern and Southern arrays and a zenith angle of 20 degrees, averaging in azimuth. 
We decided not to include the Galactic astrophysical background in our analyses. We recall that Galactic diffuse models are still highly uncertain in the TeV energy domain~\cite{2015ApJ...815L..25G, 2017PhRvL.119c1101G, 2019JCAP...12..050C} 
and, in any case, we do not expect them to play a relevant role in this particular analysis. This is so because our targets are located far away from the Galactic plane, where the Galactic diffuse emission is expected to be very subdominant. 

\begin{table}[h]
\centering
\begin{tabular}{|cccccc|}
\hline
Target & $m_\chi$ \small{[TeV]} & $T_{obs}$ \small{[h]} & (RA, DEC) \small{[deg]} & $ J \scriptsize{\left[\frac{\mathrm{GeV}^2}{ \mathrm{cm}^{5}}\right]}$ & $(E_{min},\, E_{max})$ \small{[TeV]} \\
\hline
Draco & $[0.1, 60]$ & 300 & (260.05, 57.915) & $1.42 \cdot 10^{19}$ & (0.03, $m_\chi$) \\ 
Sculptor & $[0.1, 60]$ & 300 & (15.0375, -33.7092) & $3.56 \cdot 10^{18}$ & (0.03, $m_\chi$) \\
\hline
\end{tabular}
\caption{\label{tab:simu}
Targets used in this analysis and most relevant parameters for the CTA simulations. Sources are assumed to be point-like. Right ascension (RA) and declination (DEC) refer to the target coordinates, which are slightly different from the center of the CTA field of view (we use an offset of 0.5 deg for the wobble observations; see text for details). J-factors obtained from~\cite{Bonnivard:2015xpq}. 
}
\end{table}

\subsection{Annihilation cross section values}
\label{sec:subresul}

\begin{figure}[ht]
\begin{center}
\includegraphics[width=.9\textwidth]{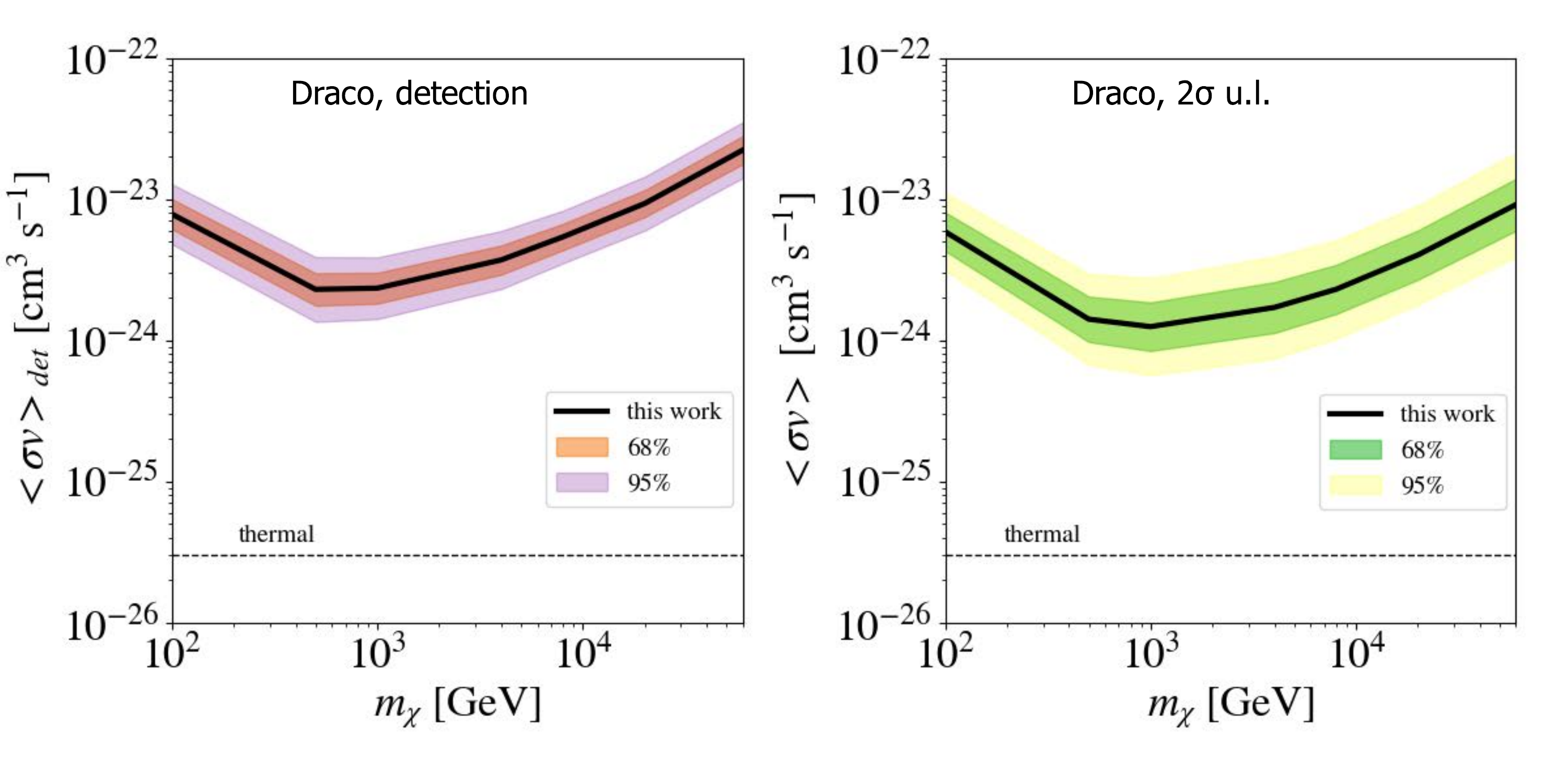}
\caption{Annihilation cross section values, as a function of the branon mass, needed for a $5\sigma$ detection with CTA (left) and for the case of setting $2\sigma$ upper limits in the absence of a branon signal (right). These results refer to 300 h of observation time of the Draco dSph by CTA-North, adopting the parameters detailed in table~\ref{tab:simu}. Coloured areas correspond to 68\% and 95\% confidence bands as obtained from 100 independent observation simulations. 
The horizontal dashed line corresponds to the thermal value of the annihilation cross section~\cite{2012PhRvD..86b3506S}. 
}
\label{fig:sigmavsdr}
\end{center}
\end{figure}

\begin{figure}[ht]
\begin{center}
\includegraphics[width=.9\textwidth]{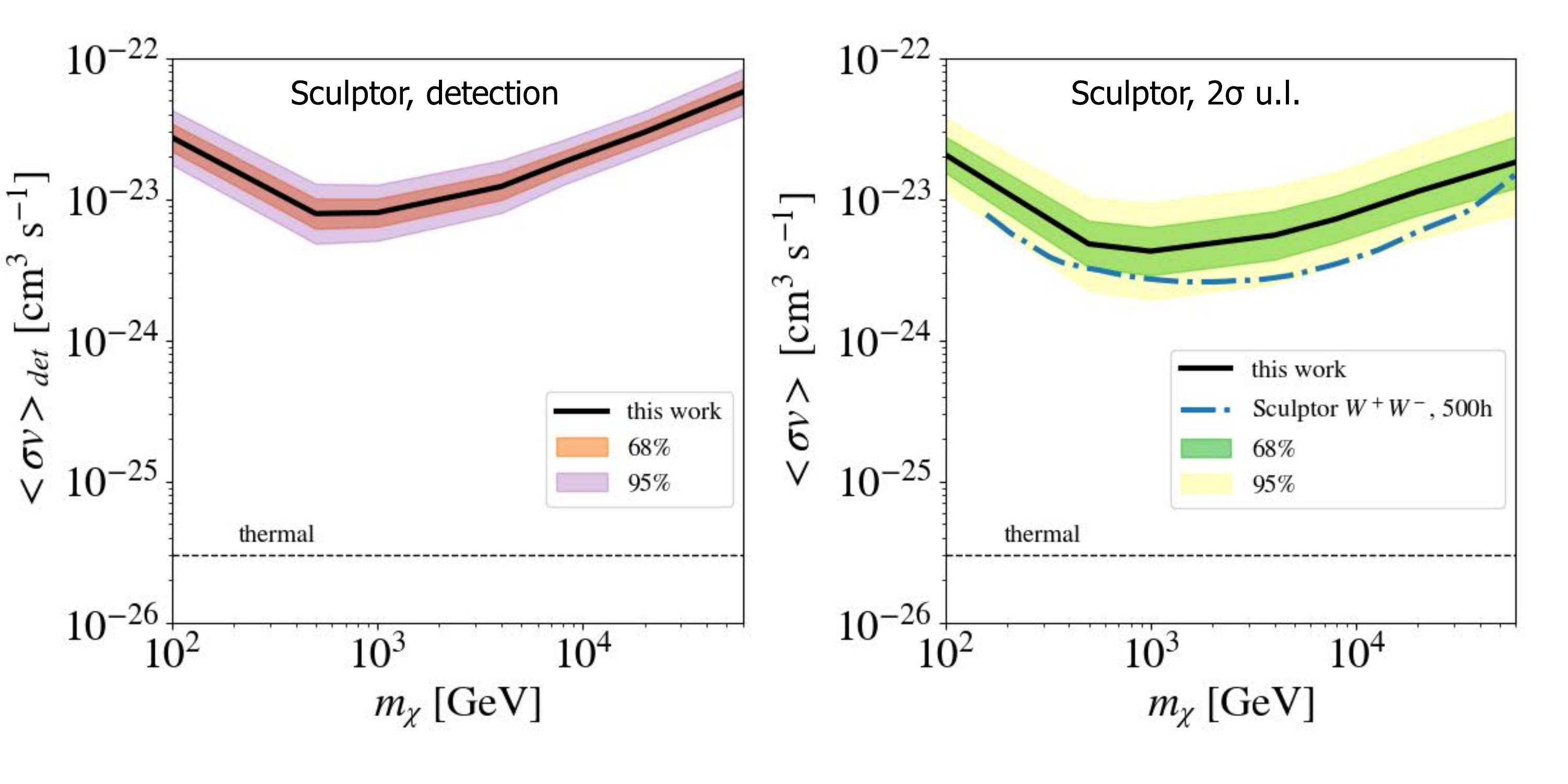}
\caption{ 
Annihilation cross section values, as a function of the branon mass, needed for a $5\sigma$ detection with CTA (left) and for the case of setting $2\sigma$ upper limits in the absence of a branon signal (right). These results refer to 300 h of observation time of the Sculptor dSph by CTA-South, adopting the parameters detailed in table~\ref{tab:simu}. Coloured areas correspond to 68\% and 95\% confidence bands as obtained from 100 independent observation simulations. The horizontal dashed line corresponds to the thermal value of the annihilation cross section~\cite{2012PhRvD..86b3506S}. In the right panel, we also show the upper limits predicted for this same object (dash-dotted blue line) in case of WIMPs annihilating to $W^+W^-$ and 500 hours of observation taken from~\cite{Carr:2015hta}. }
\label{fig:sigmavssc}
\end{center}
\end{figure}

In this section, we present and discuss the results we obtained after having implemented the simulation chain in section~\ref{sec:met} for Draco and Sculptor and those parameters in table~\ref{tab:simu}. 
Figures~\ref{fig:sigmavsdr} and~\ref{fig:sigmavssc} show, for Draco and Sculptor respectively, the annihilation cross section values that would be needed for detection (left panels) and those  corresponding to 95\% c.l. upper limits (right panels). 
In each case, we performed one hundred simulations and employed different simulation seeds, which allowed us to obtain the $1\sigma$ and $2\sigma$ confidence bands shown in these figures.\footnote{Since CTA is not built yet, we use different seeds to simulate different CTA observations, and obtain mean values and corresponding standard deviations. We performed a study of the convergence of the latter as the number of simulations increase; the results of these tests are included in appendix~\ref{sec:app-converg}.} 
 From the left panels, one can conclude that it is very unlikely that CTA could detect thermal branons in these objects. Indeed, in case of an hypothetical detection, we would be two orders of magnitude further away from the thermal relic cross section. A similar situation can be seen in the right panels of figures~\ref{fig:sigmavsdr} and~\ref{fig:sigmavssc}: CTA will not be able to rule out any mass for thermal branons, the exclusion limits lying around 1.5 orders of magnitude, at best, above the thermal cross section value. 

In the right panel of figure \ref{fig:sigmavssc}, we also compare the obtained upper limits with results shown in~\cite{Carr:2015hta} for Sculptor and the $W^+W^-$ channel. We note that this channel is the dominant one for branons above the kinematic threshold for W production (see right panel of figure~\ref{fig:bryf}), which motivates this comparison. The differences found are reasonable and can be due to the fact of using different IRFs (ours being the most recent) and observation times (300 vs 500$\,h$). 

Once again, we must note that, in the computation of the branon annihilation spectra, the electroweak corrections have been computed at leading order, which might not be precise enough above $\sim$10 TeV~\cite{Ciafaloni:2010ti}.  This is the main reason to not show results above 60 TeV in figures~\ref{fig:sigmavsdr} and~\ref{fig:sigmavssc}. In any case, we remind that CTA sensitivity at these extremely high energies is statistic limited, thus worsening as the energy increases simply due to the properties of the studied branon spectra.

In the case the branon is assumed to be a thermal relic, we recall that the tension of the brane is fully determined by the mass of the branon (figure~\ref{fig:bryf}, left panel). In this scenario, it is then possible to translate limits to the annihilation cross section as a function of the branon mass into limits to the branon tension as a function of the branon mass. To do so, we translate the obtained 95\% upper limits on the annihilation cross section into $2\sigma$ exclusion regions in the tension-mass parameter space first introduced in figure \ref{fig:bryf} in the following way: 

\begin{equation} 
f = \left( \frac{\hbar^2 c^3}{\expval{\sigma v}} \sum_i d_{0,i} \right)^{1/8},
\end{equation}

\noindent where we sum over all annihilation channels $i$, $\hbar$ is the reduced Planck constant, $c$ is the speed of light, and the channel-dependent constants, $d_{0,i}$, proportional to the leading order component of the annihilation cross section into channel $i$, are described in appendix~\ref{sec:appcs}.

\begin{figure}[ht]
\begin{center}
\includegraphics[width=.7\textwidth]{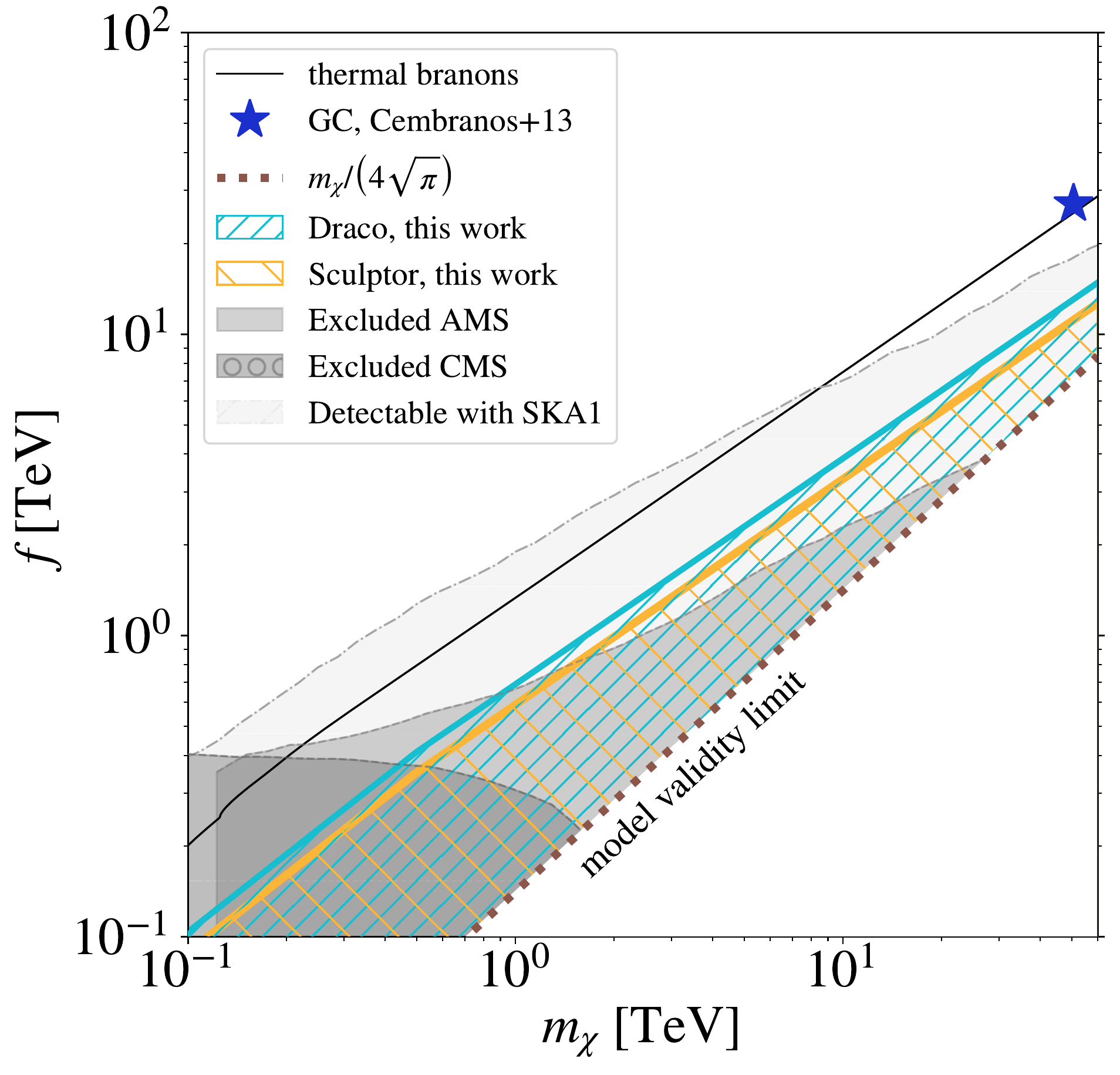}
\caption{Branon tension as a function of its mass. The thick dotted brown line at the very bottom shows the limit of validity for the model in the $f(m_\chi)$ parameter space.
The black line --already shown in figure~\ref{fig:bryf}-- shows the relation between tension and branon mass if these were thermal relics. Orange and blue areas correspond to regions of the parameter space that would be excluded at $2\sigma$ c.l. by the upper limits obtained in the right panels of figures~\ref{fig:sigmavsdr} and~\ref{fig:sigmavssc} for Draco and Sculptor, respectively. 
Dark gray regions are those that have been already excluded by CMS and AMS, while the light gray region refers to prospects for SKA from synchrotron radio emission expected from DM annihilation in Draco~\cite{2020PDU....2700448C}. The dark blue star, also taken from ref.~\cite{2020PDU....2700448C}, refers to the branon model that would provide an alternative explanation to the observation of very high energy gamma-rays at the Galactic center (GC) with HESS~\cite{2013ehep.confE.103C}. }
\label{fig:resbrane}
\end{center}
\end{figure}

The result of this study is illustrated in figure~\ref{fig:resbrane}, where it is shown that CTA will exclude a large region of the mass-tension parameter space, $f(m_\chi)$, enclosing approximately mass values between 0.1 and 60 TeV, and tension values up to 10 TeV for the largest masses. In the same figure, we also compare with results obtained in ref.~\cite{2020PDU....2700448C} from the Antimatter Spectrometer, AMS~\cite{2008NIMPA.588..227B}, and the Compact Muon Solenoid, CMS~\cite{2008JInst...3S8004C}, as well as prospects for the future Square Kilometre Array, SKA~\cite{2006AN....327..387G}. As it can be seen, all of these exclusion regions are nicely complementary, our results expanding the excluded region for TeV masses with respect to the limits from current experiments.\footnote{In figure~\ref{fig:resbrane} we show, for the case of AMS, the most conservative limits, i.e. those obtained within the hypothesis that branons represent --together with a minimum background component--  the main contribution to the AMS data. These constraints appear to be almost independent from both the adopted DM density profile for the Galaxy and the diffusion model. On the other hand, by assuming that branons represent only a small contribution to the measured positron fraction --whose main contribution can be explained by pulsars--, the corresponding AMS exclusion limits would be significantly more stringent; yet, they would largely depend on the DM distribution profile in the Galaxy and diffusion model. In the most stringent case, by assuming a Navarro-Frenk-White density profile with a maximum diffusion model, these limits reach $f=10$ TeV at $m_\chi=70$ TeV \cite{Cembranos:2017eie}.}
Yet, in the future, SKA will be the most sensitive instrument in this parameter space, encompassing all of the mentioned excluded regions, and excluding thermal branons with masses up to a few TeV. Indeed, radio telescopes possess a superb sensitivity with respect to other detectors in this context (e.g. figure 8 in \cite{2020PDU....2700448C}). Yet, it must be noted that DM searches via radio interferometry, are affected by important systematic uncertainties. In particular, it is challenging to accurately model  the involved magnetic fields as well as the physics of diffusion of charged particles, both key for an accurate estimate of the synchrotron radio emission induced by DM annihilation events (see \cite{2020PDU....2700448C} and references therein). 

\section{Discussion and conclusions}\label{sec:conclu}

In this work, we have studied the sensitivity of the future Cherenkov Telescope Array to branons, the DM candidate originated by the effective field theory of a brane-world scenario. This candidate is of particular interest for gamma-ray telescopes sensitive to the TeV energy domain, due to the possibility of the model to generate a heavy and thermally produced DM candidate with a mass up to 100 TeV. In the limit of thermal branons, the branching ratios into different annihilation channels depend only on the mass of the particle. The multi-TeV branon DM candidate was first considered as an alternative explanation for the very-high-energy gamma-ray emission from the Galactic center as measured by the previous generation of gamma-ray telescopes; in particular, a branon with a mass of 50 TeV combined with a power-law background component seemed to fit well the gamma-ray spectral cut-off observed by H.E.S.S. in the inner 10 parsecs of the Galactic center~\cite{PhysRevD.86.103506, Cembranos:2013fya, Gammaldi:2019mel, Gammaldi:2016uhg}. 

To address the CTA sensitivity to branons, we decided to focus on two of the best astrophysical targets traditionally identified in gamma-ray DM searches, namely the Draco and Sculptor dSphs, located in the Northern and Southern Hemispheres, respectively. We have simulated 100 CTA observations for each of these objects by adopting the branons' annihilation spectra discussed in section~\ref{sec:anibra} for different branon masses, and analyzed our simulations with the help of \texttt{ctools} and following the procedure described in section~\ref{sec:submet}. We explored two scenarios in our analysis: detection and upper limits to the annihilation cross section in case of no signal. In the former, we computed the required averaged annihilation cross sections required for a CTA detection, while in the latter we first obtained the corresponding 95\% c.l. upper limits to the annihilation flux for different branon masses, and then translated this information into 95\% c.l. upper limits to the annihilation cross section as well. 

We found that the cross section values needed to have a $5\sigma$ detection of branons in Draco and Sculptor are around two orders of magnitude, at best, larger than the thermal relic cross section value, i.e., several times $10^{24}\, \mathrm{cm}^3\, \mathrm{s}^{-1}$, as illustrated in the left panels of figures~\ref{fig:sigmavsdr} and~\ref{fig:sigmavssc}. Therefore, we conclude that it is unlikely that we will be able to detect thermal branons with CTA using Draco and Sculptor. With these results in mind, we proceeded and computed 95\% confidence level upper limits to branon annihilation cross section for both dSphs. This is shown in the right panels of figures~\ref{fig:sigmavsdr} and~\ref{fig:sigmavssc}, respectively. Despite we are not able to reach the thermal relic cross section --our constraints being around $1.5-2$ orders of magnitude above the thermal relic cross section-- these results are competitive and comparable to CTA predictions for generic WIMPs~\cite{Carr:2015hta}. As such, our constraints improve the best ones obtained by current IACTs for WIMP annihilation in dSphs by around a factor $\sim 5-10$ (see e.g.~\cite{2014JCAP...02..008A, 2014PhRvD..90k2012A, 2018JCAP...03..009A, 2020arXiv200305260M, 2017PhRvD..95h2001A}), showing once again the enhanced capabilities of CTA to test TeV DM candidate models.  

We note that the above results could be improved by taking into account other, non-considered effects. For instance, the enhancement in the central density associated with a BH-induced DM-spike may boost the gamma-ray flux \cite{Gondolo:1999ef, Gammaldi:2016uhg}, and thus improve our constraints by a factor of a few up to $10^3$, depending on the BH mass (typically, between $10^3 - 10^6 M_\odot$ in dSphs) and the DM density profile (the maximum boost is associated to cusp-like profiles) \cite{Gammaldi:2016uhg, Gonzalez-Morales:2014eaa}. On the other hand, we note that the benchmark Sommerfeld correction \cite{Lattanzi:2008qa} does not apply to branons. In fact, they do not possess electroweak charge: branons couple directly to SM particles without any mediator. 

Though CTA will not be able to test thermal branons by means of Draco and Sculptor observations, this will allow us to exclude a significant portion of the brane tension versus branon mass parameter space, $f(m_\chi)$, ranging from 0.1 to 60 TeV in branon mass and up to $\sim$10 TeV of the brane tension, as shown in figure~\ref{fig:resbrane}. 
In a more general context, we note that CTA will be enlarging the region of the parameter space that has already been excluded by AMS and CMS by other means, especially for branon masses above 1 TeV. CTA will also probe a large fraction of the exclusion region achievable by SKA using radio observations, this way providing valuable complementary information.

In the future, we may explore other astrophysical targets to understand the full capabilities of CTA to constrain branon models, such as the Galactic center, galaxy clusters or even other, potentially more promising dwarfs than those studied here. A stacking analysis following the observations of several dwarfs would help increasing the sensitivity in our study as well. 
As a general remark from this first work on CTA and branons, we conclude that CTA possess a superb potential to pursue this kind of DM searches and, in particular, to test well-motivated TeV DM candidates like the one here discussed. In this enterprise, CTA will also offer the necessary complementarity with other current and future instruments that may test branon models by means of different messengers, approaches or frequencies, such as AMS, CMS and SKA.

\appendix

\acknowledgments

This work was conducted in the context of the Dark Matter and Exotic Physics Working Group of the CTA Consortium and received valuable comments from Vitor de Souza and Daniel Kerszberg.

 The authors would like to thank Luigi Tibaldo for his valuable help with \texttt{ctools}-related questions, as well as Christopher Eckner and Eirik S{\ae}ther Hatlen for useful discussions on the statistical aspects of this work, and J.A.R. Cembranos for useful discussions on brane-world physics.
 
 The work of AAS, VG and MASC was supported by the Spanish Agencia Estatal de Investigaci\'on through the grants PGC2018-095161-B-I00 and IFT Centro de Excelencia Severo Ochoa SEV-2016-0597, the {\it Atracci\'on de Talento} contract no. 2016-T1/TIC-1542 granted by the Comunidad de Madrid in Spain, and the MultiDark Consolider Network FPA2017-90566-REDC. The work of AAS was also supported by the Spanish Ministry of Science and Innovation through the grant FPI-UAM 2018.
 
 DN acknowledges support from the former Spanish Ministry of Economy, Industry, and Competitiveness / European Regional Development Fund grant FPA2015-73913-JIN and the MultiDark Consolider Network FPA2017-90566-REDC.

VG's contribution to this work has been supported by Juan de la Cierva-Formaci\'on FJCI-2016-29213 grant.

This research made use of ctools, a community-developed analysis package for IACT data. ctools is based on GammaLib, a community-developed toolbox for the high-level analysis of astronomical gamma-ray data. Also, this research made use of Python, along with community-developed or maintained software packages, including IPython \cite{Ipython_paper}, Matplotlib \cite{Matplotlib_paper}, NumPy \cite{Numpy_paper} and SciPy \cite{2020SciPy-NMeth}. 
Numerical computations were made using the Hydra cluster at the Instituto de F\'isica Te\'orica (Universidad Aut\'onoma de Madrid) and the computational resources of the High Energy Physics Group at Universidad Complutense de Madrid. This work made use of NASA's Astrophysics Data System for bibliographic information.

This research has made use of the CTA instrument response functions provided by the CTA Consortium and Observatory, see \url{http://www.cta-observatory.org/science/cta-performance/} (version prod3b-v2) for more details.

\section{Branon annihilation cross sections} \label{sec:appcs}

We have made use of the leading order component of the annihilation cross sections calculated for branons to get the parameter tension $f$ in section~\ref{sec:subresul}. As per~\cite{Cembranos:2003fu}, in natural units:

\begin{equation}
\expval{\sigma v}  =\sum_{n=0}^{\infty}c_nx^{-n}\Rightarrow \expval{\sigma v}f^8  \approx \sum_i d_{0,i},
\end{equation}

where $x=M/T$, $d_{0,i} =  f^8 c_{0,i}$ and $c_{0,i}$ depends on the particle $i$. 
For Dirac fermions, $\psi$:
\begin{equation}
    c_{0,\psi} = \frac{1}{16\pi^2 f^8} m_\chi^2 m_\psi^2 (m_\chi^2 - m_\psi^2) \sqrt{1-\frac{m_\psi^2}{m_\chi^2}}
\end{equation}

For a massive gauge field, $Z$:
\begin{equation}
    c_{0,Z} = \frac{m_\chi^2 \sqrt{1-\frac{m_Z^2}{m_\chi^2}} (4m_\chi^4 - 4m_\chi^2 m_Z^2 + 3m_Z^4)}{64\pi^2 f^8} 
\end{equation}

And for a complex scalar field, $\Phi$:
\begin{equation}
    c_{0,\Phi} = \frac{m_\chi^2 \sqrt{1-\frac{m_\Phi^2}{m_\chi^2}} (2m_\chi^2 - m_\Phi^2)^2}{32\pi^2 f^8} 
\end{equation}

In all cases, $m_\chi$ is the branon mass.

\section{Cross section uncertainty bands} \label{sec:app-converg}

\begin{figure}[ht]
\begin{center}
\includegraphics[width=.9\textwidth]{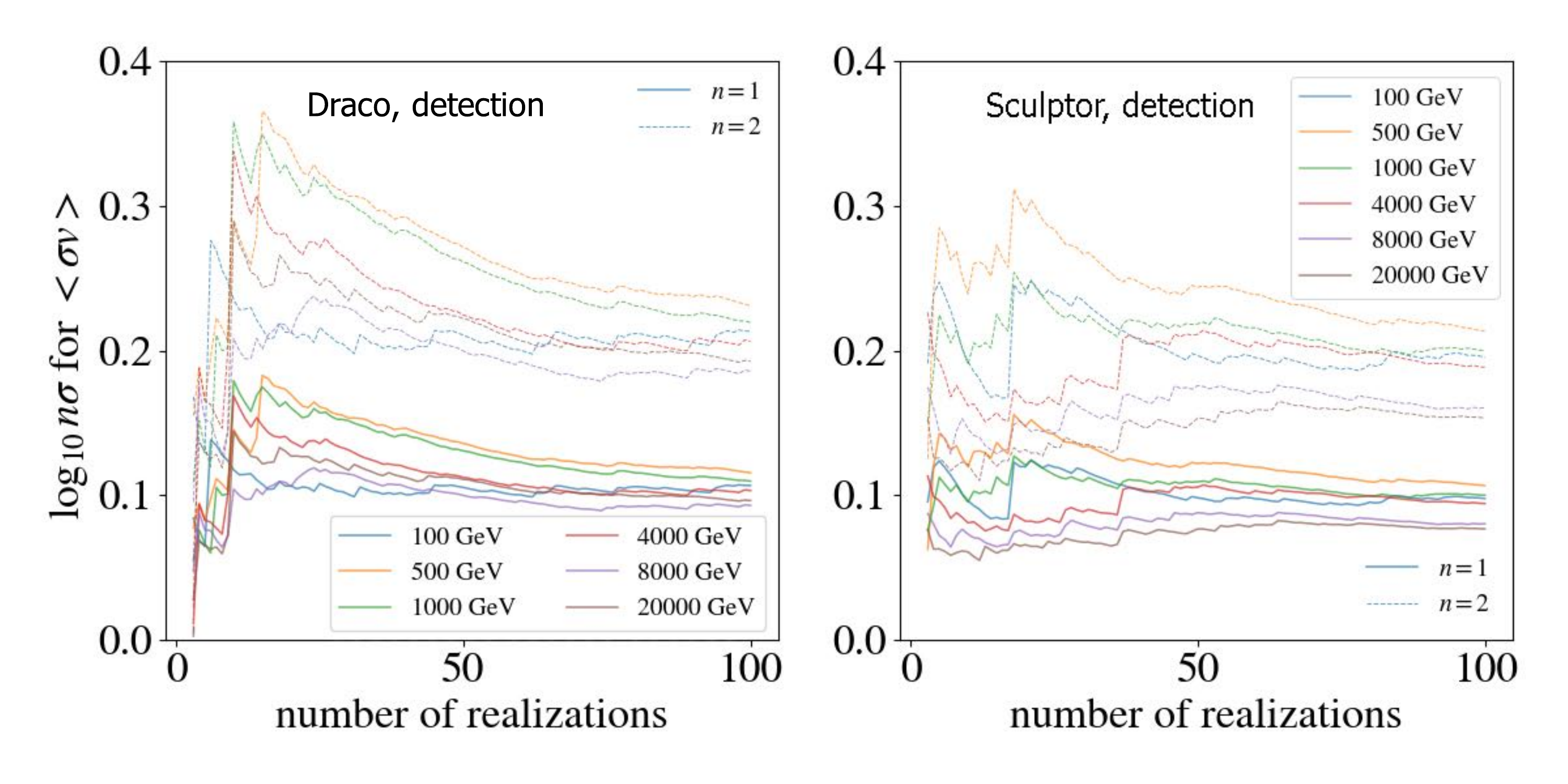}
\includegraphics[width=.9\textwidth]{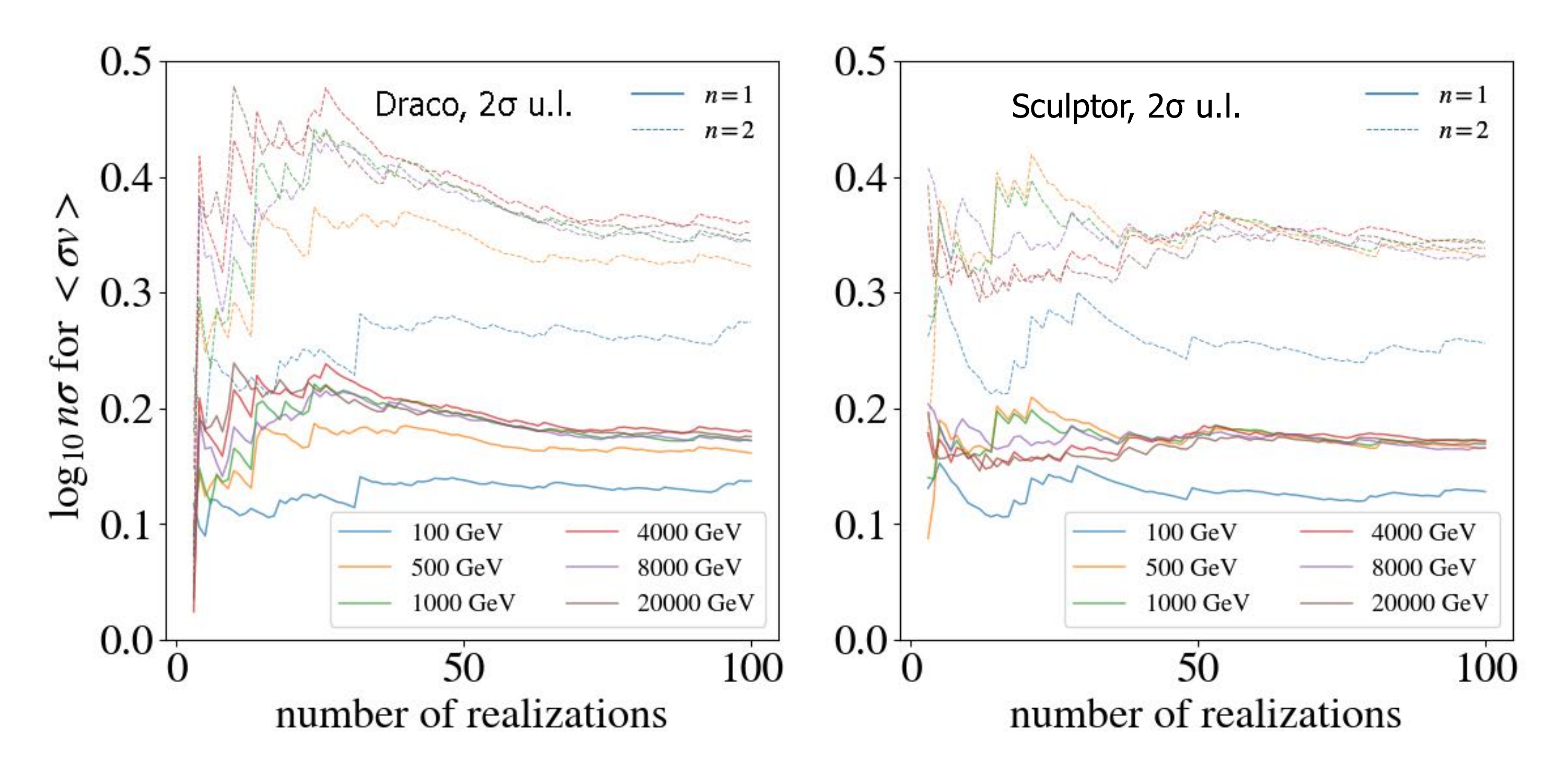}
\caption{ 
Study of convergence of the $1\sigma$ and $2\sigma$ values on the annihilation cross section both for Draco (left) and Sculptor (right), and for different branon masses. Top and bottom panels correspond, respectively, to the case of detection and upper limits. Note that, implicitly, this study assumes a lognormal distribution of the data in all cases. The $1\sigma$ and $2\sigma$ uncertainty bands that were shown in figures~\ref{fig:sigmavsdr} and~\ref{fig:sigmavssc} correspond to the $1\sigma$ and $2\sigma$ values reached after 100 simulations in these plots.
}
\label{fig:sigmavserr}
\end{center}
\end{figure}

In this appendix, we summarize our studies on the convergence of the width of the $1\sigma$ and $2\sigma$ uncertainty bands shown in the annihilation cross section figures of section~\ref{sec:subresul}. More specifically, we studied their behaviour as a function of the number of performed simulations of a CTA observation. We recall that each of these simulations start with a different seed in \texttt{ctobssim}. This exercise allows us to easily obtain a qualitative estimate of the number of realizations that are needed in order to reach a reasonable convergence of results. This is shown in figure~\ref{fig:sigmavserr}, where we observe that $1\sigma$ and $2\sigma$ values converge after some tens of iterations, both for the case of studying detection and for the case of setting upper limits. We conclude that a reasonable compromise is $\sim$50 simulations in all cases. Yet, to err on the side of caution, the results presented in this work come from sets of 100 simulations.

\bibliographystyle{JHEP.bst}
\bibliography{References.bib}

\end{document}